# Temporal and Dose Kinetics of Tunnel Relaxation of Non-Equilibrium Near-Interfacial Charged Defects in Insulators

Gennady I. Zebrev, Maxim G. Drosdetsky

*Abstract* — **This paper is devoted mainly to mathematical aspects of modeling and simulation of tunnel relaxation of non-equilibrium charged oxide traps located at/near the interface insulator - conductive channel, for instance in irradiated MOS devices. The generic form of the tunnel annealing response function was derived from the rate equation for the charged defect buildup and annealing as a linear superposition of the responses of different defects with different time constants. Using this linear response function, a number of important practical problems are analyzed and discussed. Combined tunnel and thermal or RICN annealing, power-like temporal relaxation after a single ion strike into the gate oxide, are described in context of general approach.**

## I. Introduction

The devices of MOS technologies are the basis of modern microelectronics and, therefore, an adequate radiation response prediction for conditions of space low-dose-rate irradiation based on laboratory tests is very important task. [1, 2, 3]. The process of radiation degradation in the MOS devices consists of several stages [4, 5, 6]. The first stage is radiation-induced creation of the electron – hole pairs in the insulator followed by their separation under action of the oxide electric field $E_{ox}$. The efficiency of the electron-hole separation is characterized by a dimensionless electric field dependent function $\eta(E_{ox})$. This function is generally less than unity due to recombination processes. Radiation-induced electrons, having relatively large mobility, are swept out the oxide under action of the electric fields (~ $10^5 - 10^6$ V/cm in thin gate oxides). A fraction $F_{ot}$ (typically $\ll 1$) of the radiation-induced holes are trapped at the deep-energy oxide defects, located within a few nanometers near the Si-SiO$_2$ interface (see Fig. 1). These positively charged defects, identified as E'-centers [7, 8, 9], are relatively stable, and most of them have energies typically beneath the valence band of silicon. Due to the proximity of the Si substrate the near-interfacial defects are capable to exchange charge with underlying silicon via electron tunneling [10, 11].

Since the energy levels of the defects are located below the Fermi level in the Si substrate, relaxation process always proceeds in the direction of positive trapped charge compensation (permanent anneal).

Depending on the spatial location of the charged traps their characteristic recharging times may extend in very wide range. Charged defects, located in the oxide bulk at distances greater than a few nanometers from the Si-SiO$_2$ interface may recharge only at very large times, and, thus, are usually treated as a fixed (i.e., gate bias-independent) positive oxide charge ($Q_{ot}$). The defects with the energy levels, located opposite to the silicon bandgap, are capable to reversibly exchange the carriers with the Si substrate, depending on the Fermi level position[12, 13, 14]. Such defects are traditionally referred to as interface traps, border traps, or switching states [15, 16, 17, 18].

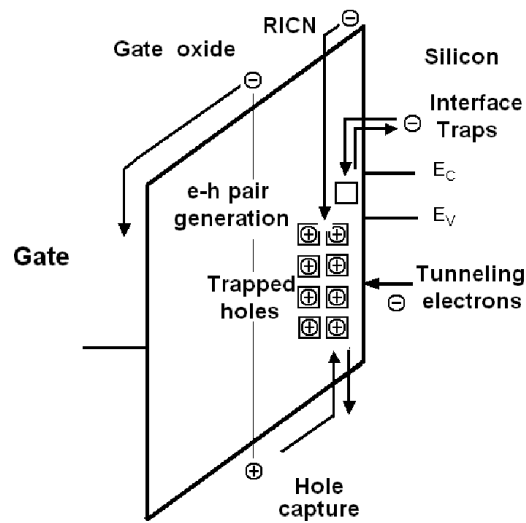

Fig. 1. Basic radiation-induced processes in gate oxides of MOS devices.

Neutralization of positively charged defects in SiO$_2$, located in energy below the Si valence band edge, occurs at a rate that is approximately independent of temperature and linear with logarithmic time [19, 20, 21, 22]. This remarkable property is a direct consequence of an exponentially wide range of the tunnel relaxation times. The tunneling of electrons as mechanism of logarithmic dependence has been examined by several investigators [23, 24, 25, 26].

The linear theory in a form of temporal convolution integral with the empirical forms of logarithmic response function was used for description of long-term tunnel annealing at different dose rates [27, 28]. The main objective of this paper is to provide a consistent mathematical structure for description of delayed temporal kinetics of the tunnel relaxation at the different dose-rate profiles, based on an exact solution of the linear kinetic equation.

## II. MATHEMATICAL FORMALISM AND APPLICATIONS

### A. Rate equation for buildup and tunnel relaxation

Radiation hardness of the silicon – oxide insulator interface of can be parameterized by the effective width of the oxygen vacancy precursors for the deep hole traps $\ell$, which is as thin as a few nanometers, and the hole trapping efficiency, which can be parameterized as follows

$$F_{ot} = \Sigma_p N_{VO} \ell, \qquad (1)$$

where $N_{VO}$ is the oxygen vacancy bulk density, and $\Sigma_p$ is, in general, the electric-field-dependent deep hole capture cross-section ($\Sigma_p \sim 10^{-15}$ cm$^2$ [29]). The experimentally observed range of $F_{ot} \sim 10^{-3} – 10^{-2}$ suggests that $N_{VO}$ has to be of order $10^{19} – 10^{20}$ cm$^{-3}$.

For definiteness, we will consider here the n-channel MOSFET at positive gate biases ($V_G > 0$). The hole capture rate is proportional to radiation-induced hole flux. Then the rate equation for the trapped hole charge bulk density $q_{ot}$ reads

$$\frac{dq_{ot}(x)}{dt} = \frac{F_{ot}}{\ell} q A_d P - \frac{q_{ot}(x)}{\tau(x)}, \qquad (2)$$

where $P$ is the dose rate, $A_d = K_g d_{ox} \eta(E_{ox})$, $q$ is the electron charge, $\eta(E_{ox})$ is the electric field dependent electron-hole charge yield, $d_{ox}$ is the oxide thickness, $K_g \cong 8 \times 10^{12}$ cm$^3$rad(SiO$_2$)$^{-1}$ is the electron-hole pair generation rate constant in SiO$_2$ [4]. For brevity, the rate equation (2) is written in this section in a simplified form, which implies the inequality $q_{ot} \ll qN_{VO}$. More general case will be discussed in Appendix B.

Tunnel relaxation times are spread on many orders [30]

$$\tau(x) = t_{min} \exp\left(\frac{x}{\lambda}\right), \qquad (3)$$

$t_{min}$ is the cutoff tunneling time, $\lambda$ is the minimal tunneling length ($\leq 0.1$ nm).

### B. Generic form of tunnel relaxation response function

The kinetic equation (2) is considered in this paper as a linear one, i.e. all the model parameters in (2) are assumed to be independent on $q_{ot}$ (or, the same, on the dose). This approximation is justified only at relatively low doses. The high dose conditions generally require a numerical self-consistent consideration [31]. The linearity condition makes it easy to get the exact analytical solution of Eq. 2. Integrating this solution over $x$, one can obtain a relationship for the surface density of positive trapped charged defect $Q_{ot}$ as a convolution of temporal dose-rate pre-history

$$Q_{ot}(t) = qF_{ot} A_d \int_0^t \Re_{tun}(t-t')P(t')dt'. \qquad (4)$$

Here, $\Re_{tun}(t-t')$ is the linear response function calculated as result of integration over uniform distribution of the trapped holes in a form

$$\Re_{tun}(t-t') = \frac{\lambda}{\ell}\left(E_1\left(\frac{t-t'}{t_{max}}\right) - E_1\left(\frac{t-t'}{t_{min}}\right)\right), \qquad (5)$$

where $E_1(y)$ is the integral exponential function [32], $t_{max} = t_{min} \exp(\ell/\lambda)$ is the maximum tunneling time. The cutoff time parameter $t_{min}$ has to be formally determined by the tunnel response of the fastest trap. In practice, $t_{min}$ is bounded below by the hole transport temporal scales (typically within $10^{-5}$ s). Fig. 2 shows a generic form of the linear response function of the tunnel annealing $\Re_{tun}(t)$.

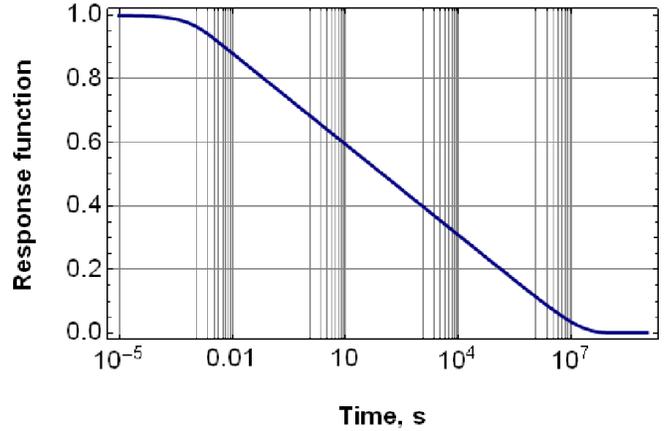

Fig. 2. Simulated with Eq.5 the response function of tunnel relaxation annealing calculated $t_{min} = 10^{-5}$ s, $t_{max} = 3 \times 10^7$ s. Such temporal range of recharging times corresponds to the trapped hole layer thickness $\Delta \ell = \lambda \log(t_{max}/t_{min}) \cong 24 \lambda$.

Response function is equal to unity at $t = t'$ ($\Re(0) = 1$) and has strictly zero value at $t < t'$ (causality property). Radiation response at any dose-rate profile $P(t)$ can be calculated using generic response function $\Re_{tun}(t)$ with a use of a convolution integral in Eq. 4.

### C. Pulse irradiation

Typical durations of the irradiation pulse (e.g., in the electron accelerators) are rather small. It allows one to formally express the dose rate profile in a form of the Dirac delta function $P(t) = D\delta(t)$, where $D$ is a total dose per a pulse. Then, the temporal kinetics of the tunnel relaxation is given by a simple function of time, elapsed after an irradiation pulse

$$Q_{ot}(t) = qF_{ot}A_d D \frac{\lambda}{\ell}\left(E_1\left(\frac{t}{t_{max}}\right) - E_1\left(\frac{t}{t_{min}}\right)\right). \qquad (6)$$

If $\ell \ll d_{ox}$, then oxide-trapped threshold voltage shift component is given by

$$\Delta V_{ot}(t) \cong -Q_{ot}(t)/C_{ox}, \qquad (7)$$

where $C_{ox}$ is the specific oxide capacitance.

The integral exponent function $E_1(x)$ has at $x \ll 1$ an asymptotic form in terms of natural logarithm $E_1(x) \cong -\gamma - \ln x$, where $\gamma \cong 0.577$ is the Euler's constant [32]. On the condition that $t_{max} \gg t \gg t_{min}$ we have an asymptotic form

$$|\Delta V_{ot}(t)| \cong |\Delta V_{ot}^{max}|\left[1 - \frac{\lambda}{\ell}\log\left(\frac{t}{t_{min}}\right)\right]. \quad (8)$$

This gives the well-known logarithmic dependence which is a characteristic feature of tunnel mechanism of trapped charge annealing [2, 33]. Neglecting of Euler's constant in (8) is expressed only in insignificant redefinition of the ill-defined constant $t_{min}$.

The slope of the annealing curve in logarithmic time axis is given by a derivative with respect to the logarithm of time

$$\frac{1}{|\Delta V_{ot}^{max}|}\frac{\partial|\Delta V_{ot}(t)|}{\partial \ln t} = -\frac{\lambda}{\ell}. \quad (9)$$

Thus, it would seem that the dimensionless ratio $\lambda/\ell$ can be experimentally determined immediately from the slope of curve $|\Delta V_{ot}(t)|$ normalized to its maximum value. The difficulty is that the voltage shift is very rapidly decreasing function at small temporal scales and its exact value immediately after the pulse $|\Delta V_{ot}^{max}|$ remains, as a rule, unknown. This problem requires more careful consideration.

*D. Parameter Renormalization*

Analysis shows that the problem of characterizing of the annealing rate is closely associated with characterization of the hole trapping efficiency factor $F_{ot}$. Suppose we have a "true" value of $F_{ot}$ defined without tunneling relaxation processes. Logarithmic dependence in (8) has a remarkable property that stems from a very fast temporal change of function at the small time scales ($t \geq t_{min}$) and very slow behavior at large elapsed times ($t \gg t_{min}$). This leads to occurrence of dependence of measured results on typical temporal scales of measurements. For example, the slope of the experimental curve $|\Delta V_{ot}(t)|$, normalized to the maximum value, turns out to be dependent on the choice of the time of the first measurement $t_1$ which typically obeys the inequality $t_1 \gg t_{min}$. Particularly, the experimentally measured effective $F_{ot}^*$ is a decreasing function of $t_1$, since a significant part of the trapped positive charge has at the time of measurement already compensated by tunneling electrons. This circumstance is formally expressed as a remarkable property of Eq.6 which can be represented equivalently in a two-fold manner

$$C_{ox}|\Delta V_{ot}(t)| =$$
$$= qF_{ot}A_d D\left(1 - \frac{\lambda}{\ell}\ln\left(\frac{t}{t_{min}}\right)\right) = qF_{ot}^*A_d D\left(1 - \frac{\lambda}{\ell^*}\ln\left(\frac{t}{t_1}\right)\right), \quad (10)$$

where the renormalized values $F_{ot}^*$ and $\ell^*$ are introduced as follows

$$F_{ot}^* = F_{ot}\left[1 - \frac{\lambda}{\ell}\ln\left(\frac{t_1}{t_{min}}\right)\right], \quad \ell^* = \ell\left[1 - \frac{\lambda}{\ell}\ln\left(\frac{t_1}{t_{min}}\right)\right]. \quad (11)$$

Of course, equations (10) and (11) are valid only providing that $t \geq t_1 > t_{min}$. A concrete choice of $t_1$ is dependent particularly on characteristic dose rate magnitudes, varying in practice from a fraction of seconds for the pulsed irradiation case to hours or, even, days for irradiation at low dose rates.

Notice, that the parameter renormalization leaves invariant a combination of parameter which is dependent on the deep hole trap creation cross-section, and on concentration of oxygen vacancies

$$\frac{F_{ot}}{\ell} = \frac{F_{ot}^*}{\ell^*} = \Sigma_p N_{VO}. \quad (12)$$

Dependence on renormalization procedure (i.e. on arbitrary choice of the time of first measurement $t_1$) means that initial conditions play important role in parameter extraction. Particularly, as can be seen from (10) and (11), the choice of $t_1$ impacts the free term characterizing initial condition, leaving unchanged an absolute magnitude of the logarithmic rate, expressed, for instance, in mV/decade

$$C_{ox}|\Delta V_{ot}(t)| = qF_{ot}^*A_d D - qA_d D(\Sigma_p N_{VO}\lambda)\ln\left(\frac{t}{t_1}\right), \quad (12)$$

$$C_{ox}\frac{\partial|\Delta V_{ot}(t)|}{\partial \ln t} = -qA_d D(\Sigma_p N_{VO}\lambda). \quad (13)$$

Recall, that $D$ in this section is a constant dose per an irradiation pulse. Nevertheless, the equation (13) shows that the slope of logarithmic time dependence which can be determined immediately from experimental curve, does not provide information about effective thickness of trapped holes $\ell$.

Procedure of renormalization excludes from consideration typically unobservable and poorly determined values of maximum and minimum tunneling times. At the same time, the renormalization introduces the dependence of description on the experimental conditions of measurements such as the time of first measurement. Such dependence is an inevitable consequence of the specific type of logarithmic relaxation in tunneling processes.

*E. Steady-state irradiation*

Let us consider the case of steady-state irradiation with a constant dose rate $P$ during exposure time $t_{irr}$ ($D(t) = Pt$). Radiation response is expressed in this case as follows

$$Q_{ot}(t) = qF_{ot}A_d P T_B(t), \quad (14a)$$

where the temporal buildup function $T_B(t)$ is defined as a result of exact analytic integration in Eq. 4

$$T_B(t) = \frac{\lambda}{\ell}\int_0^t \left[E_1\left(\frac{t-t'}{t_{max}}\right) - E_1\left(\frac{t-t'}{t_{min}}\right)\right]dt' =$$

$$= \frac{\lambda}{\ell}t\left(E_1\left(\frac{t}{t_{max}}\right) - E_1\left(\frac{t}{t_{min}}\right)\right) +$$

$$+ \frac{\lambda}{\ell}t_{max}\left(1-\exp\left(-\frac{t}{t_{max}}\right)\right) - \frac{\lambda}{\ell}t_{min}\left(1-\exp\left(-\frac{t}{t_{min}}\right)\right).$$
(14b)

For large $\ell/\lambda$, the temporal buildup function is almost identical to the current time of irradiation $T_B(t) \le t$.

On the condition that $t_{max} \gg t \gg t_{min}$, Eq. 14 leads to the following asymptotic form

$$Q_{ot}(D,P) \cong qF_{ot}A_d D\left(1-\frac{\lambda}{\ell}\left(\ln\left(\frac{D}{Pt_{min}}\right)-1\right)\right) =$$
$$qF_{ot}^*A_d D\left(1-\frac{\lambda}{\ell^*}\ln\left(\frac{D}{Pt_1}\right)\right),$$
(15)

where the renormalized parameters read as

$$F_{ot}^* = F_{ot}\left[1-\frac{\lambda}{\ell}\left(\ln\left(\frac{t_1}{t_{min}}\right)-1\right)\right], \quad \ell^* = \ell\left[1-\frac{\lambda}{\ell}\left(\ln\left(\frac{t_1}{t_{min}}\right)-1\right)\right].$$
(16)

The magnitude of $F_{ot}^*$ will vary depending on the choice of the first measurement. This choice depends eventually on the steady-state dose-rate value, and it can be strongly different for dose rates in its typical ranges from mrad(Si)/s to 100 rad(Si)/s. In practice, this circumstance leads to significant reduction of the slope of dose curves at low dose rates. Fig. 3 shows the simulated dose response functions at steady-state irradiation, calculated at different dose rates taking into account the tunnel annealing with Eq. 14b.

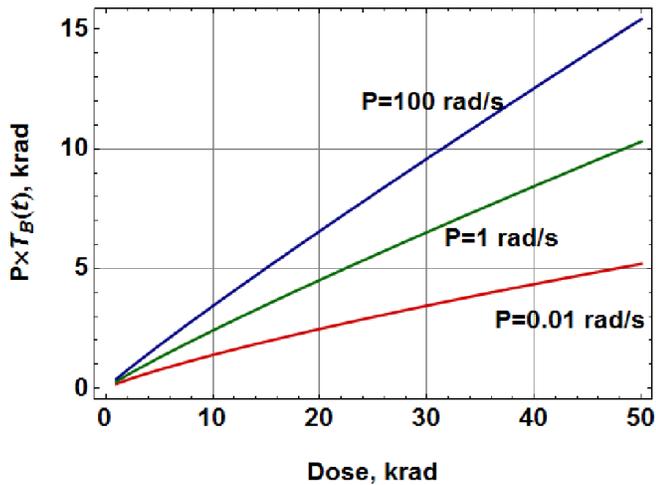

Fig. 3. "Effective dose" function defined as $P \times T_B(D/P)$ at three different dose rates calculated with Eq. 14b as function of total ionizing dose. Used parameters are $\lambda/\ell = 1/45$, $\lambda = 0.1$ nm, $t_{min} = 10^{-4}$ s.

To avoid misunderstanding, it should be emphasized that the tunnel annealing is a typical time-dependent effect, and it does not belong to the class of the true dose-rate effects such as the ELDRS [34]. The difference in slope of the quasi-linear dose curves at different dose rates in Fig. 3 is due solely to the fact that the dose curves are plotted on different time scales. Note, that the slopes of the dose curves on a linear scale in Fig. 3 are approximately proportional to renormalized values of $F_{ot}$.

As can be seen in Fig. 3, the process of oxide charge buildup under a fixed dose is essentially suppressed with decreasing dose rate because of the simultaneous tunnel annealing. This circumstance is rather essential in low-dose-rate environment operation, for example, in space. Under very large time scales $t_{irr} > t_{max}$ the tunneling relaxation rate equals to the hole trapping rate and the oxide charge would saturate formally on a level

$$Q_{ot}^{sat} \cong qF_{ot}A_d \frac{\lambda}{\ell}Pt_{max}.$$
(17)

In practice, the maximum tunneling times $t_{max}$ are too big (at least, much greater than a few years). For instance, the results, presented in [33], provide an opportunity to estimate $t_{max}$ as large as $\sim 10^9$ s (~30 years), corresponding approximately to $\ell \sim 32\,\lambda$. Therefore, radiation-induced oxide charge buildup at large doses are limited by other mechanisms, such as the thermal anneal [35], the Radiation-Induced Charge Neutralization (RICN) effects [36], or, the eventual saturation of the charge trapping due to depletion of the oxygen precursor density. These aspects will be described and discussed in Appendices A and B.

*F. Tunnel relaxation after irradiation*

In this section we will consider a process of tunnel annealing of positive trapped charge as function of time $\Delta t$, elapsed after ending of steady-state irradiation with a duration $t_{irr}$. A general formula gives for conditions $t_{max} \gg t$, $t_{irr} \gg t_{min}$ the following asymptotic relationship

$$\frac{|\Delta V_{ot}(\Delta t, t_{irr})|}{|\Delta V_{ot}(t_{irr})|} = 1 - \frac{\lambda}{\ell^*}\left(\ln\left(1+\frac{\Delta t}{t_{irr}}\right) + \frac{\Delta t}{t_{irr}}\ln\left(1+\frac{t_{irr}}{\Delta t}\right) - 1\right).$$
(18a)

where

$$\ell^* = \ell - \lambda\ln(t_{irr}/t_{min}), \quad F_{ot}^* = F_{ot} - \Sigma_p N_{VO}\lambda\ln(t_{irr}/t_{min}).$$
(18b)

As in the preceding sections, we operate here with renormalized values when $|\Delta V_{ot}(t_{irr})|$ corresponds to a measured value at time $\Delta t = 0$ of threshold voltage shift, that already reduced by tunnel annealing. At large times $\Delta t \gg t_{irr}$

$$C_{ox}|\Delta V_{ot}(\Delta t, t_{irr})| \cong qF_{ot}^*A_d D\left(1-\frac{\lambda}{\ell^*}\ln\left(\frac{\Delta t}{t_{irr}}\right)\right)$$
(19)

The asymptotic relations (18-19) are useful for a qualitative understanding of the time dependence shapes. Exact and explicit (though cumbersome) design equations, convenient for numerical calculations are presented in Appendix D.

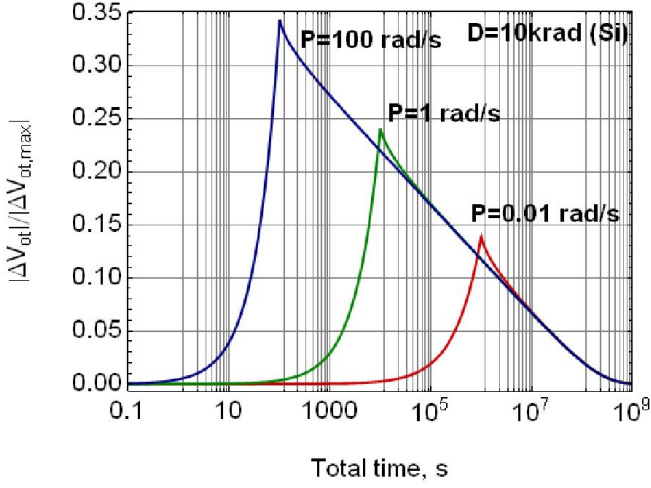

Fig. 4. Normalized voltage shift as functions of total (irradiation + anneal) time for different dose rates 10 krad; Used parameters are $\lambda / \ell = 1/45$, $\lambda = 0.1$ nm, $t_{min} = 10^{-4}$ s.

Calculations in Fig. 4 show the temporal dependencies of degradation during and after irradiation with different dose rates. During irradiation, all the curves are different, while some times after the end of irradiation ($\sim t_{irr}$), the curves, as functions of the total elapsed time, practically coincide. Such remarkable experimental fact has been observed by many authors [27, 28, 33, 37].

### G. Power-like annealing after pulse irradiation

Typically, logarithmic view of temporal dependence can be clearly detected only in the measurements carried out on several orders of magnitude in time. However, there are situations where the logarithmic kinetics manifests itself on the linear time scale. For instance, the subthreshold current of MOSFETs is a parameter which is very sensitive to fine temporal kinetics of buildup and annealing of the oxide-trapped charge.

Most clearly this can be seen under pulsed irradiation. Then, the subthreshold leakage current is expressed as follows

$$I_{leak}(t) \propto \exp\left(\frac{|\Delta V_{ot}(t)|}{m\varphi_T}\right), \quad (20)$$

where $\varphi_T = k_B T/q$ is the thermal potential, $m$ is non-ideality factor (typically $1 < m < 2$), and $|\Delta V_{ot}(t)|$ is a function of the time, elapsed after irradiation pulse. Using Eq. 10, one can get

$$\frac{I_{leak}(t > t_1)}{I_{leak}(t_1)} \cong \exp\left(-\alpha \ln\left(\frac{t}{t_1}\right)\right) = \left(\frac{t_1}{t}\right)^\alpha, \quad (21)$$

where a dimensionless parameter, defined as,

$$\alpha = \Sigma_p N_{VO} \frac{q K_g d_{ox} \eta(E_{ox}) D}{C_{ox} m \varphi_T}, \quad (22)$$

is dependent on the pulse dose and on the oxide's parameters. Power-like fast annealing of subthreshold current in power MOSFETs after strikes of the single heavy ions was experimentally observed in [38].

### H. Relation with true dose-rate effects

Tunnel relaxation is a time-dependent effect. Common feature of time-dependent annealing processes is suppression of degradation with decreasing dose rate. In contrast, the true dose rate effects, such as ELDRS, exhibit opposite tendency. According to [39], the ELDRS effect is due to decreasing of the charge yield with dose rate increase, caused by the excess trap-assisted recombination in insulators. Thus, the competition between opposite types of dose rate dependencies in the true-dose-rate and the time-dependent processes is formally expressed in competition of two factors

$$|\Delta V_{ot}| \propto \eta(E_{ox}, P) F_{ot} D \left(1 - \frac{\lambda}{\ell^*} \ln\left(\frac{D}{P t_1}\right)\right). \quad (23)$$

Similar competition between the tunnel annealing and the RICN effect was discussed in [40].

### III. CONCLUSION

Based on analytical solution of differential kinetic equation, a consistent description of logarithmic-like tunnel relaxation as a convolution integral over a dose-rate profile with an exact linear response impulse function has been developed. All parameters of this approach have clear physical meaning and can be extracted from experimental dose dependences. It is shown that model parameters are renormalized at different dose-rate profiles. Measured parameters are, in fact, the renormalized ones. In particular, due to such renormalization, the measured slope of the quasi-linear dose curves are decreasing function of the dose rates. This circumstance necessitates the determination of the invariant parameters to recalculate the results to a different dose rate.

### APPENDICES

### A. Combined impact of thermal and tunnel anneal

In contrast to tunnel-assisted relaxation, thermal anneal is position-independent and often can be approximated by a discrete spectrum of the time constants. Supposing for brevity a single time constant of thermal $\tau_a = \tau_{a0} \exp(\varepsilon_a / k_B T)$ ($\varepsilon_a$ is an activation energy, $\tau_{a0}$ is a temperature-independent time constant), the rate equation reads

$$\frac{dq_{ot}(x)}{dt} = \frac{F_{ot}}{\ell} q A_d P - q_{ot}(x)\left(\frac{1}{\tau(x)} + \frac{1}{\tau_a}\right). \quad (A1)$$

By doing so, as was done above, we have found the response function of combined tunnel and thermal relaxation in a multiplicative form

$$\Re(t-t') = \Re_{tun}(t-t') \exp\left(-\frac{t-t'}{\tau_a}\right). \quad (A3)$$

To obtain a saturation value of degradation at constant-dose-rate steady-state irradiation, we have to set the upper limit integration in Eq. 4 to infinity. Then we have

$$Q_{ot}^{SAT} = qF_{ot}A_d P \int_0^\infty \Re(t-t')dt' \cong$$
$$\cong qF_{ot}A_d P \frac{\lambda}{\ell} \int_0^\infty E_1\left(\frac{t-t'}{t_{max}}\right) \exp\left(-\frac{t-t'}{\tau_a}\right) dt'. \quad (A4)$$

Using a standard integral, we find an exact result

$$Q_{ot}^{SAT} = \frac{\lambda}{\ell} qF_{ot}A_d P\tau_a \ln\left(1 + \frac{t_{max}}{\tau_a}\right). \quad (A5)$$

On the typical condition $t_{max} \gg \tau_a$, Eq. A5 reads as follows

$$Q_{ot}^{SAT} \cong qF_{ot}A_d P\tau_a \left(1 - \frac{\lambda}{\ell} \ln\frac{\tau_a}{t_{min}}\right). \quad (A6)$$

Without tunnel relaxation (formally, at $\lambda \to 0$) we would have saturation on a level $Q_{ot}^{SAT} \cong qF_{ot}A_d P\tau_a$. The tunneling processes reduce the maximum trapped charge as well as at a standard renormalization procedure (see Sec. II-D)

$$F_{ot} \to F_{ot}\left(1 - \frac{\lambda}{\ell} \ln\frac{\tau_a}{t_{min}}\right). \quad (A7)$$

At sufficiently low temperatures, when $t_{max} \ll \tau_a$ and thermal annealing is suppressed, we would have a result, which is fully consistent with Eq.17. Notice that the saturation charge in Eq.17 is divergent at $\lambda \to 0$ due to lack of any annealing mechanisms. In practice, there exist other mechanisms of the trapped charge restriction such as the RICN or depletion of the precursor density. Both these restriction mechanisms correspond to the time constant which turns out to be inversely proportional to the dose rate (see App.B).

### B. Influence of precursor trap density depletion

There is another fundamental mechanism of the charge trapping saturation associated with depletion of the precursor trap density $N_{VO}$. This mechanism can be accounted for by generalizing of the rate equation (A1) as follows

$$\frac{dq_{ot}(x)}{dt} = \Sigma_p \left(qN_{VO} - q_{ot}(x)\right) A_d P - q_{ot}(x)\left(\frac{1}{\tau(x)} + \frac{1}{\tau_a}\right) =$$
$$= q\Sigma_p N_{VO} A_d P - q_{ot}(x)\left(\frac{1}{\tau(x)} + \frac{1}{\tau_a} + \frac{1}{\tau_V}\right) = \quad (B1)$$
$$= \frac{F_{ot}}{\ell} qA_d P - q_{ot}(x)\left(\frac{1}{\tau(x)} + \frac{1}{\tau_S}\right),$$

where

$$\tau_V^{-1} = \Sigma_p \eta(E_{ox}) K_g d_{ox} P = \Sigma_p A_d P, \quad (B3)$$

$$\tau_S = \left(\tau_a^{-1} + \tau_V^{-1}\right)^{-1} = \frac{\tau_a}{1 + \Sigma_p A_d P\tau_a}. \quad (B4)$$

Equation (B1) is equivalent to Eq. A1. Similarly, we find the saturation charge as follows

$$Q_{ot}^{SAT} = q\frac{\lambda}{\ell} F_{ot}A_d P\tau_S \ln\left(1 + \frac{t_{max}}{\tau_S}\right). \quad (B5)$$

At $t_{max} \ll \tau_S$ the Eq.17 is reproduced again, while at $t_{max} \gg \tau_S$ we have

$$Q_{ot}^{SAT} = qF_{ot}A_d P\tau_S\left(1 - \frac{\lambda}{\ell} \ln\frac{\tau_S}{t_{max}}\right). \quad (B6)$$

Of course, the number of the charged traps in any case can not be greater than the density of its precursors (i.e. the oxygen vacancies). Particularly, at $\tau_V \ll \tau_a < t_{max}$, we have

$$Q_{ot}^{SAT} < qF_{ot}A_d P\tau_V = \frac{qF_{ot}}{\Sigma_p} = qN_{VO}\ell. \quad (B7)$$

If $t_{max} \gg \tau_S$, the saturation of this type occurs formally on the condition

$$P\tau_a > \left(\Sigma_p \eta(E_{ox}) K_g d_{ox}\right)^{-1} \sim \frac{1}{\eta(E_{ox})}\left(\frac{\mu m}{d_{ox}}\right) \text{ Mrad}, \quad (B8)$$

i.e. at the relatively high dose rates, thick oxides and low temperatures. Typical value of $\tau_a$ at room temperatures is of order $\sim 10^7$ s, hence the condition (B8) corresponds to $P > (10\text{-}100)$ rad/s at $d_{ox} = 10$ nm. Formally, the trapped charge saturation is determined by the minimum of the times $\tau_V, \tau_a$ and $t_{max}$.

### C. Continuum activation energy distribution

Discrete spectrum of the defects relaxation times leads to exponential temporal kinetics of annealing processes. Logarithmic view of temporal kinetics in tunneling annealing arises as a superposition of exponential temporal curves with a specific exponential scatter of time constants as in Eq. 3. Thermal annealing time constants also can have exponential scatter in magnitude due to a spread in annealing activation energy

$$\tau_a(\varepsilon_a) = \tau_{a0} \exp\left(\frac{\varepsilon_a}{k_B T}\right) \quad (C1)$$

Then, assuming for simplicity a uniform distribution of activation energy in a range from $\varepsilon_1$ to $\varepsilon_1 + \Delta\varepsilon$, we obtain

$$\Re_{therm}(t-t') =$$
$$= \int_{\varepsilon_1}^{\varepsilon_1 + \Delta\varepsilon} \frac{d\varepsilon}{\Delta\varepsilon} \exp\left(-\frac{t-t'}{\tau_a(\varepsilon)}\right) = \frac{k_B T}{\Delta\varepsilon}\left(E_1\left(\frac{t-t'}{\tau_{max}}\right) - E_1\left(\frac{t-t'}{\tau_1}\right)\right), \quad (C2)$$

where

$$\Delta\varepsilon = k_B T \ln\left(\frac{\tau_{max}}{\tau_1}\right). \quad (C3)$$

Thus, the response function of combined tunneling and thermal annealing of the defects with uniform spatial and energy distribution is deduced from (A1) as

$$\Re(t-t') = \Re_{tun}(t-t')\Re_{therm}(t-t') \quad (C4)$$

Mathematical structure $\Re_{tun}(t-t')$ and $\Re_{therm}(t-t')$ for the traps, uniformly distributed in position and energy, is equivalent each other up to a substitution $k_B T/\Delta\varepsilon \rightleftarrows \lambda/\ell$. In particular, provided the inequality $t_{max}, \tau_{max} \gg t \gg t_{min}, \tau$ is valid, the combined response function in (B4) has a form

$$\Re(t) \cong \left(1 - \frac{\lambda}{\ell}\ln\left(\frac{t}{t_{min}}\right)\right) \times \left(1 - \frac{k_B T}{\Delta\varepsilon}\ln\left(\frac{t}{\tau_1}\right)\right). \quad (C5)$$

Thermal annealing is also linear with logarithmic time, but is accelerated with increasing temperature [41, 42]. The formula (C5) illustrates the well-known concepts of the tunneling front and the thermal front of relaxation, which move independently into the spatial and energy depth of the oxide [26]. Weak non-uniformity in the distributions does not significantly change the situation qualitatively, and only leads to a change in the slope of quasi-logarithmic time dependence.

*D. Annealing after irradiation*

Let us consider constant-dose-rate irradiation with duration $t_{irr}$, followed by annealing in the same conditions at $\Delta t \equiv t - t_{irr} > 0$. Solution of Eq. 2 for a constant trapped hole charge bulk density is given by

$$q_{ot}(x) = \frac{F_{ot}}{\ell} q A_d P \tau(x) \left(1 - \exp\left(-\frac{t_{irr}}{\tau(x)}\right)\right) \exp\left(-\frac{t - t_{irr}}{\tau(x)}\right). \quad (D1)$$

Integrating in a standard way (A1) over uniform spatial distribution of charged we found surface charge density as function of time

$$Q_{ot} = q F_{ot} A_d P \frac{\lambda}{\ell} T_A(t_{irr}, \Delta t) \quad (D2)$$

where the "anneal" temporal response function is given as follows

$$T_A(t_{irr}, \Delta t) = \int_0^\ell \tau(x) \left(1 - \exp\left(-\frac{t_{irr}}{\tau(x)}\right)\right) \exp\left(-\frac{t - t_{irr}}{\tau(x)}\right) \frac{dx}{\lambda} =$$
$$= t_{irr} \left(E_1\left(\frac{t_{irr}}{t_{max}}\right) - E_1\left(\frac{t_{irr}}{t_{min}}\right)\right) - \Delta t \left(E_1\left(\frac{\Delta t}{t_{max}}\right) - E_1\left(\frac{\Delta t}{t_{min}}\right)\right) +$$
$$+ t_{max} \left(\exp\left(-\frac{\Delta t}{t_{max}}\right) - \exp\left(-\frac{t_{irr}}{t_{max}}\right)\right) +$$
$$+ t_{min} \left(\exp\left(-\frac{t_{irr}}{t_{min}}\right) - \exp\left(-\frac{\Delta t}{t_{min}}\right)\right) \cong$$
$$\cong t_{irr} - \frac{\lambda}{\ell} \left[t_{irr} \ln\left(\frac{t_{irr} + \Delta t}{t_{min}}\right) + \Delta t \ln\left(\frac{t_{irr} + \Delta t}{\Delta t}\right)\right]$$
(D3)

Fig. 4 shows simulation results calculated with an exact form of the "buildup" (14) and the "anneal" (D3) temporal response functions ($T_B(t)$ and $T_A(t_{irr}, \Delta t)$ respectively). Notice, that $T_B(t_{irr}) = T_A(t_{irr}, 0)$.

ACKNOWLEDGMENT

The authors would like to thank R. G. Useinov, who initiated this work many years ago, for discussion.